\documentclass[10pt]{article}
\usepackage{amsmath}
\usepackage{graphicx}
\usepackage{amsfonts}
\usepackage{amssymb}

\def\tit#1{\begin{centering}\Large\bf #1 \\[5mm]\end{centering}}
\def\aut#1{\centerline{#1}}
\def\add#1{\begin{centering}\it #1 \\[5mm]\end{centering}}

\begin{document}


\tit{High nuclear polarization of $^3$He at low and high pressure
by metastability exchange optical pumping at 1.5~Tesla}

\aut{M.~Abboud, A.~Sinatra, X.~Ma\^\i tre$^*$, G.~Tastevin, P.-J.~Nacher}
\add{marie.abboud@lkb.ens.fr,alice.sinatra@lkb.ens.fr\\
 Laboratoire Kastler Brossel - Ecole Normale Sup\'{e}rieure,
24 rue Lhomond, 75005 Paris, France \footnote
{Laboratoire Kastler Brossel is a unit\'e de recherche de l'Ecole
Normale Sup\'erieure et de l'Universit\'e Pierre et Marie Curie,
associ\'ee au CNRS (UMR 8552).}\\
$^*$ U2R2M, Universit\'e Paris-Sud and CIERM - H\^opital de Bic\^etre,
94275 Le Kremlin-Bic\^etre Cedex, France \footnote
{U2R2M (Unit\'e de
Recherche en R\'esonance Magn\'etique M\'edicale) is a unit\'e de
recherche de l'Universit\'e Paris-Sud, associ\'ee au CNRS (UMR
8081)}
}

\textbf{Abstract} {Metastability exchange optical pumping of
helium-3 is performed in a strong magnetic field of $1.5$~T.
The achieved nuclear
polarizations, between $80\%$ at $1.33$~mbar and $25\%$ at $67$~mbar,
show a substantial improvement at high pressures with respect to
standard low-field optical pumping.
The specific mechanisms of metastability exchange optical pumping at high field
are investigated, advantages and intrinsic limitations are discussed.
From a practical point of view, these results open the way to 
alternative technological solutions for polarized helium-3 applications 
and in particular for magnetic resonance imaging of human lungs.}

\section{Introduction}
A gas of ground state $^{3}$He atoms in which a high degree of 
nuclear polarization is achieved offers an incredibly rich
playground in various fields of science, from statistical or
nuclear physics to biophysics and medicine~\cite{Becker}.
Depending on the targeted
application, the degree of nuclear polarization, the sample density, or
the production rate of polarized atoms should be optimized. 
A recent application, which may have an important impact on the
diagnosis of pulmonary diseases, is polarized gas magnetic resonance
imaging (MRI)~\cite{Moller}.
Clinical studies to demonstrate the relevance of this new
tool are under way in Europe and in the United States.
Yet, if a wide expertise exists in MRI to adapt the existing imaging
techniques to the case of polarized gases, the gas preparation
remains a critical stage to be transferred from physics laboratories
to hospitals. Two methods are presently used to polarize $^{3}$He:
spin-exchange with optically pumped
alkali atoms~\cite{Happer} and pure-He metastability exchange optical 
pumping (MEOP)~\cite{Colegrove}. In standard conditions, MEOP is performed
at low pressures~($1$~mbar) in a guiding magnetic field of the order
of~$1$~mT. Circularly polarized light at $1083$~nm, corresponding
to the $2^3$S-$2^3$P transition of $^3$He, is used to transfer angular
momentum to the atoms and nuclear polarization is created by hyperfine
coupling in the metastable $2^3$S state. 
Through metastability exchange collisions, nuclear polarization builds
up in the ground state. The steady-state nuclear polarization obtained by
MEOP in standard conditions rapidly decreases if the pressure
of the sample exceeds a few mbar (see below, Fig.\ref{fig:bilan}-a) 
\cite{Crampton, Gentile, Cracovie}.
Therefore a delicate polarization-preserving compression stage is necessary
for MRI where the gas should be at atmospheric pressure for
inhalation, and for all applications needing a dense sample.
In this letter, the MEOP scheme is shown to withstand large hyperfine
decoupling. A strong magnetic field of $1.5$~T 
actually improves its performances with respect to standard 
low-field optical pumping. At~$1.33$~mbar, high nuclear 
polarizations of the order of~$80\%$ are routinely obtained with 
much lighter experimental constraints. At higher pressures, the achieved
nuclear polarizations are dramatically increased compared to published
low-field results. An elementary model 
with simple rate equations is used to account for these results. 

\section{Experimental setup and methods}
\begin{figure}[htb]
\centerline{\includegraphics[width=8cm,clip=]{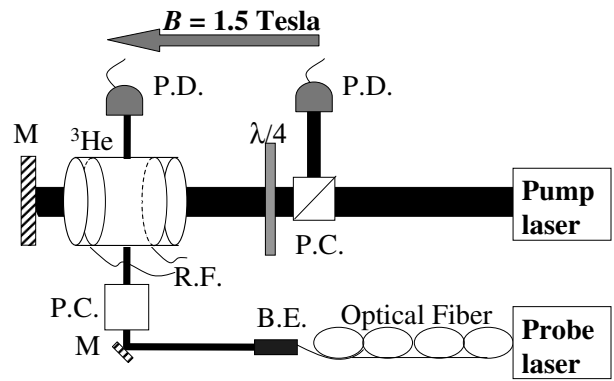}}
\caption{Experimental setup. The nuclear polarization is measured by
the absorption of a transverse probe beam. 
{\it B}: static magnetic field, P.D.: photodiode, P.C.: polarizing
beam-splitter, B.E: beam expander, $\lambda$/4: quarter wave plate, 
M: mirror, R.F.: radio-frequency to excite the discharge.}
\label{fig:setup}
\end{figure}

Experiments are performed in the bore of the $1.5$~T superconducting magnet of
a clinical MRI system. The experimental apparatus is sketched in 
Fig.\ref{fig:setup}.
The helium gas is enclosed in a sealed cylindrical Pyrex cell~(diameter~$5$~cm,
length~$5$~cm). Four cells filled with $1.33$~mbar,
$8$~mbar, $32$~mbar and $67$~mbar of pure $^{3}$He are used.
A radio-frequency discharge at $3$~MHz is sustained in the gas by external 
electrodes, leading to metastable atom densities $n_m$ 
in the $0.3$-$3\times 10^{10}$atoms/cm$^{3}$ range,
depending on the applied voltage and on the gas pressure.
The optical pumping laser is either a $50$~mW single mode laser 
diode amplified to $0.5$~W \cite{Chernikov}, or a broadband fiber laser
($1.63$~GHz FWHM) delivering $2$~W \cite{Tastevin}. 
The pump beam is back-reflected to enhance its absorption,
which is monitored on the transmitted beam with a photodiode.
At the entrance of the cell, the Gaussian transverse
intensity profile of the pump beam has a FWHM of the order of
$2$~cm. A weak probe beam from a single mode laser diode is
used to measure the nuclear polarization. 
It is linearly polarized perpendicularly to the 
magnetic field ($\sigma$ polarization).
The discharge intensity is modulated at $133$~Hz, and the probe absorption
is measured with a lock-in amplifier. Laser sources and electronics
remain several meters away from the magnet bore, in a 
low-field region.

\begin{figure}[htb]
\centerline{\includegraphics[width=13cm,clip=]{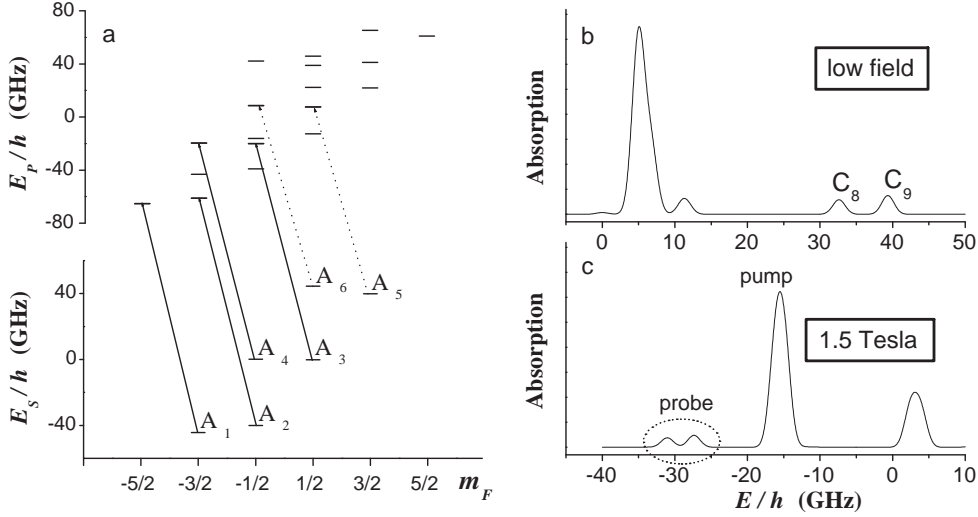}}
\caption{(a): Energies of the $^3$He sublevels at $1.5$~T for the
metastable $2^3$S state $(E_S)$ and the $2^3$P state $(E_P)$. The
transitions induced by the $\sigma^{-}$-polarized pump (solid lines)
and probe (dashed lines) are displayed. Each pump transition has a
 matrix element $T_{ij}$ close to 1\cite{Zeeman}.
(b): Absorption spectrum at low magnetic field.
(c): Absorption spectrum for $\sigma^{-}$ light at $1.5$~T. 
Vertical and horizontal scales are identical in Fig.\ref{fig:OPscheme}-b
and c. Level names A$_1$ to A$_6$, energy zeroes and spectral line positions 
are defined as in reference \cite{Zeeman}.}
\label{fig:OPscheme}
\end{figure}

At $1.5$~T, due to Zeeman splitting, the energy levels of the $2^3$S 
and $2^3$P states are spread over $80$ and $160$~GHz respectively
(Fig.\ref{fig:OPscheme}-a). 
Hyperfine decoupling in the $2^3$S state is
significant, so that the eigenstates of the Hamiltonian show only
little mixing between different eigenstates $|m_J, m_I\rangle$ of the 
decoupled spin system, where $m_J$, $m_I$, and $m_F$ denote the magnetic
quantum numbers for the electronic, nuclear, and total angular
momentum, respectively. As shown in Fig.\ref{fig:OPscheme}-a, 
the $2^3$S sublevels are arranged into three pairs of quasi-degenerate 
levels of increasing energies (A$_1$,A$_2$), (A$_3$,A$_4$), and
(A$_5$,A$_6$) that correspond respectively to $m_J$=$-1$, $0$, and $1$
in the completely decoupled limit $B\rightarrow\infty $.
For more details about the $2^3$S level structure and the analytical
expressions of eigenstates and energies, we refer the reader to 
\cite{Zeeman}.
The absorption spectra at low magnetic field and at $1.5$~T are displayed in 
Figs.\ref{fig:OPscheme}-b and c, respectively.   
In standard MEOP, very high nuclear polarizations are obtained using C$_8$ 
or C$_9$ lines \cite{Gentile, Nacher85}.
Comparable polarizations are achieved at $1.5$~T using the $\sigma^-$-strong
pump line displayed in Fig.\ref{fig:OPscheme}-c.
All the results presented in this work are obtained with this pump transition. 
The performances and efficiencies of other optical pumping 
transitions at $1.5$~T will be reported elsewhere.
The pump simultaneously addresses the four $2^3$S sublevels 
A$_1$ to A$_4$. 
Population transfer into the pair (A$_5$,A$_6$) is achieved by the 
following sequence : laser excitation, collisional 
redistribution in the $2^3$P state and spontaneous emission.
The ground state nuclear polarization $M$ is defined as 
$M$=$(n_{+}-n_{-})/(n_{+}+n_{-})$ where $n_{+}$ and $n_{-}$ 
denote populations of the $m_{I}$=$1/2$
and $m_I$=$-1/2$ nuclear spin states, respectively.
In the absence of optical pumping, metastability
exchange collisions impose a spin temperature
distribution for the $2^3$S sublevel populations, proportionally to 
$e^{\beta m_F}$ where $e^\beta$=$n_{+}$/$n_{-}$=$(1+M)/(1-M)$ 
\cite{Zeeman}.
\begin{figure}[htb]
\centerline{\includegraphics[width=13cm,clip=]{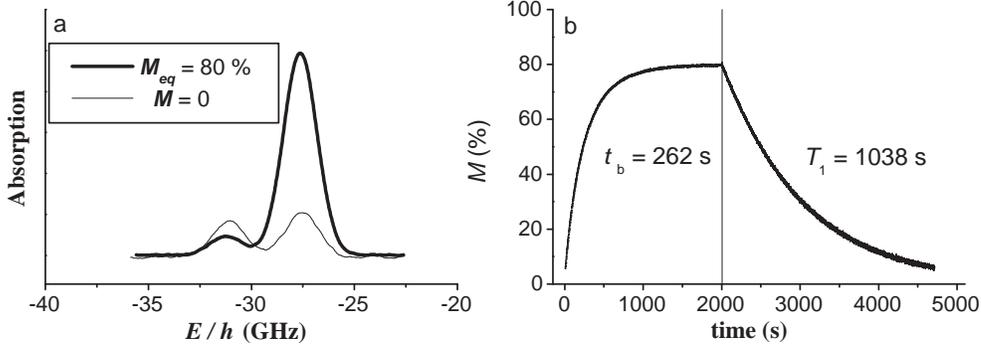}}
\caption{Examples of recorded signals in the $1.33$~mbar cell:
a-~Absorption measurements on transitions from
sublevels A$_5$ ({\it E/h}=-27.36 GHz) and A$_6$
({\it E/h}=-31.04 GHz) at thermal equilibrium 
($M\simeq$0) and at steady-state nuclear
polarization ($M$=$M_{eq}$) in an optically-pumped gas. 
b-~Polarization build-up and discharge-induced decay deduced from
changes of light absorption in sublevel A$_5$.
Pump laser is applied at time $t$=0, and turned off after 2000~s.}
\label{fig:exemple}
\end{figure}
The populations of sublevels A$_5$ and A$_6$, not
addressed by the pump, are probed to measure $M$.
Examples of probe absorption spectra for
an unpolarized and an optically-pumped steady-state situation
are shown in Fig.\ref{fig:exemple}-a. $M$ is inferred from the
relative heights of the absorption peaks.
The build-up of the polarization, as well as its decay when
the pump is turned off, are monitored by tuning the probe
laser frequency on the probe transition starting
from the A$_5$ ($m_F$=$3/2$) sublevel (Fig.\ref{fig:exemple}-b). 
These measurement procedures operate at arbitrary magnetic field
and pressure~\cite{Zeeman}.
\section{Results}
\begin{figure}[htb]
\centerline{\includegraphics[width=14cm,clip=]{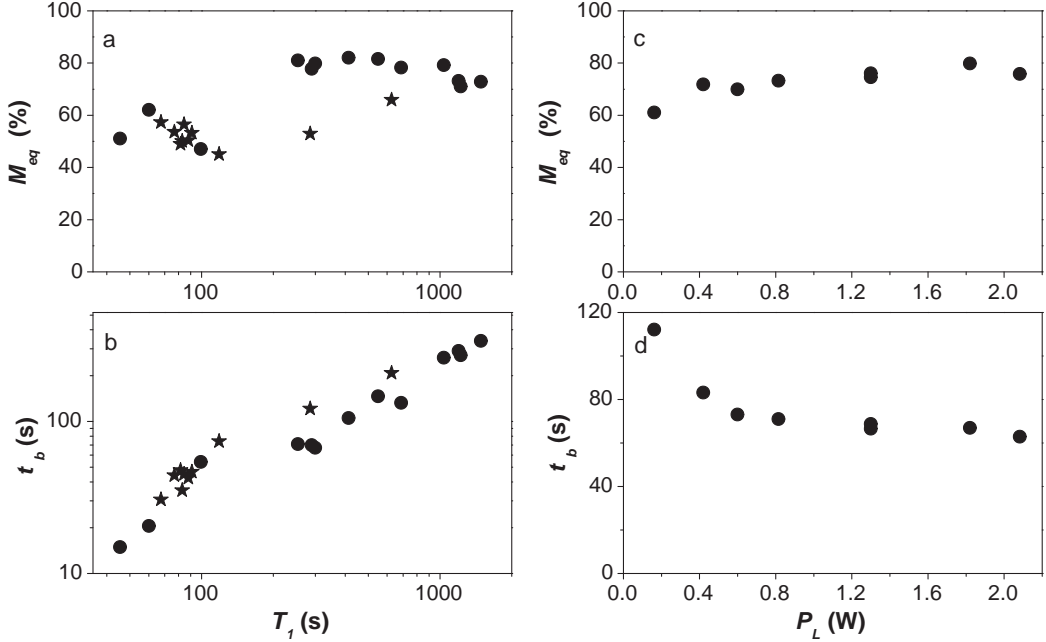}}
\caption{Results obtained at $1.5$~T in the $1.33$~mbar cell.
(a): Steady-state polarization, and (b): Polarization build-up time 
constant, as a function of the discharge-induced decay time of 
the polarization $T_1$.
Circles and stars: broadband ($2$~W) and single mode ($0.5$~W) pump
lasers, both running at full power.
(c): Steady-state polarization, and (d): Polarization build-up time
constant, as a function of incident laser power ${\cal P}_{L}$.
Data are obtained with the broadband pump laser and 
for $T_1$=$300$~s.}
\label{fig:M&tbLP}
\end{figure}

The steady-state polarization $M_{eq}$ and the polarization build-up
time constant $t_b$ in the $1.33$~mbar cell are shown in 
Figs.\ref{fig:M&tbLP}-a and b as a function of the 
discharge-induced decay time $T_1$.
Over a wide range of moderate to weak discharges 
($T_1$ ranging from $300$~s to $1500$~s), $t_b$ (ranging from
$60$ to $350$~s) is proportional to $T_1$ and the polarization
achieved with the broadband~$2$~W laser is high, about $80\%$,
independently of $T_1$. 
This behavior is specific to the high-field
optical pumping, and contrasts with the standard low-field situation
where a very weak discharge is required to obtain such large 
nuclear polarizations.
For the strongest discharges, build-up times decrease 
($t_b$ ranging from $15$ to $55$~s) and steady-state
polarizations are lower.
Fig.\ref{fig:M&tbLP}-c and d show the influence of the pump laser power
for a given discharge ($T_1$=$300$~s).
A laser power as low as $0.5$~W is sufficient for the polarization 
and the build-up time to almost reach their asymptotic values.
\begin{figure}[htb]
\centerline{\includegraphics[width=14cm,clip=]{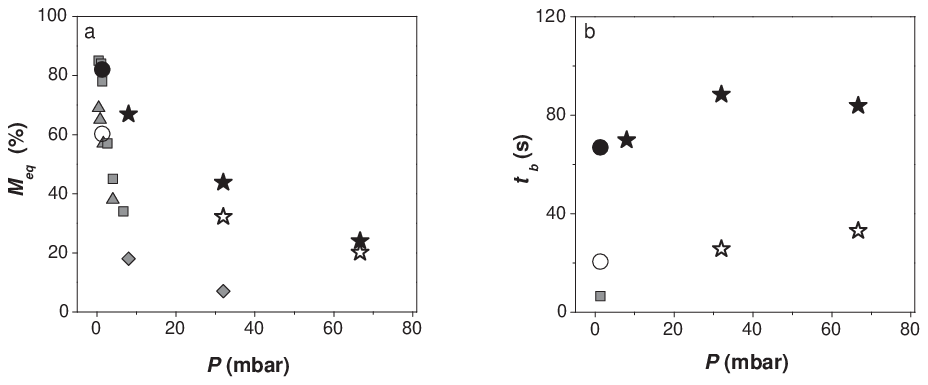}}
\caption{(a): Steady-state polarization, and (b): Polarization build-up time 
constant, as a function of $^3$He pressure $P$, 
at high and low magnetic fields.
Circles and stars are $1.5$~T data obtained with a broadband ($2$~W) and
single mode ($0.5$~W) pump lasers respectively.
Filled (open) symbols are for weak (strong) discharge :
$T_1$=$300~(60)$, $2600$, $1600~(325)$, and $1300~(700)$~s for $1.33$, 
$8$, $32$, and $67$~mbar, respectively.
Triangles, squares, and diamonds are low-field data 
published in \cite{Crampton}, \cite{Gentile}, and \cite{Cracovie}
respectively (all for weak discharges).}
\label{fig:bilan}
\end{figure}  
Similar studies of MEOP have been performed at higher pressures, where
$T_1$ between $300$ and $2600$~s are measured. 
Selected results for a weak and a strong
discharge at full laser power are shown in Fig.\ref{fig:bilan}
together with published low-field results.
The polarizations obtained at high pressures are, to our 
knowledge, record MEOP values. 
The polarization build-up times weakly depend on
$^3$He pressure, in contrast with low-field MEOP~\cite{Gentile,Stoltz}. 

\section{Discussion}
An extension of the detailed model for standard MEOP \cite{Nacher85}
to high-field conditions \cite{Zeeman} is required to compute the populations 
of all atomic sublevels.
Here, for simplicity, an elementary model is used to account
for the main observed features.
We assume that ({\it i})~atoms are fully pumped into the (A$_5$,A$_6$)
pair, and ({\it ii})~the populations of sublevels not addressed by the
pump laser are imposed by the ground state spin temperature which
only depends on $M$: $a_5$=$(1+M)/2$ and $a_6$=$(1-M)/2$.
The sublevel A$_5$ is totally oriented ($m_J$=1, $m_I$=1/2) and carries 
a nuclear angular momentum $\langle I_z\rangle$=$\hbar/2$, 
while A$_6$ has a small component on ($m_J$=0, $m_I$=1/2)
and a large component on ($m_J$=1, $m_I$=-1/2) and thus carries a
nuclear angular momentum $\langle I_z\rangle$=$\hbar(\epsilon-1)/2$
with $\epsilon$=$1\times10^{-2}$ at $1.5$~T \cite{noteEPS}.
The rate equation for $M$, resulting from
relaxation and metastability exchange, is then:
\begin{equation}
\frac{dM}{dt}=\frac{2\langle I_z\rangle/\hbar - M}{T_e} - \frac{M}{T_1}
\hspace{1cm} 
\mbox{with}
\hspace{1cm}
\langle I_z \rangle=\frac{\hbar}{2}(M+\epsilon \frac {1 - M}{2}) \, ,
\nonumber
\label{eq:dM}
\end{equation}
where $1/T_e$ is the metastability exchange
collision rate for a $^3$He atom in the ground state 
($1/T_e$=$n_m\times 1.53\times10^{-10}$cm$^3$/s), 
and $2\langle I_z \rangle/\hbar$ is the nuclear polarization in 
the $2^3$S state. One infers a steady-state polarization
$M_{eq}$=$(1+2T_e/(\epsilon T_1))^{-1}$ 
and a build-up time $t_b$=$2T_eM_{eq}/\epsilon$.
Using values of $n_m$ and $T_1$ measured in the absence 
of pumping beam, the predicted polarization is $M_{eq}$$\simeq$$1$,
at all pressures. The build-up times are in the range
$20$-$300$~s for the low pressure cell, and in the range $15$-$40$~s
for the three high pressure cells.
Although this elementary model is clearly not sufficient to predict 
$M_{eq}$, it accounts reasonably well for the observed dynamics.
Its domain of validity and accuracy are
estimated from detailed rate equations for the six $2^3$S-sublevel populations.
We find that in our experimental conditions and for the observed range of
nuclear polarization, $2\langle I_z \rangle/\hbar$ given by the simple
model differs from the exact value by a factor not exceeding 2, depending on 
$M$ and on the gas pressure. This difference results
from incomplete population transfer into (A$_5$,A$_6$) as well 
as from deviations of the order of $\epsilon$
of the populations $a_5$ and $a_6$ from their assumed
spin-temperature values.  
In spite of its simplicity, the model sheds light on
two key features: 
({\it i})~The dynamics of optical pumping at $1.5$~T is intrinsically
limited by hyperfine decoupling.
({\it ii})~The build-up time, at least in the explored range of
parameters, weakly depends on pressure 
and is affected only through changes of $n_m$ and $T_1$.

For application purposes, production rates of polarized atoms
per unit volume $R_a$=$P M_{eq}/t_b$ are
compared to published results for standard MEOP 
conditions and similar sealed cells in Table~\ref{table:R_a}.
\begin{table}[h]
\caption{Steady-state polarizations $M_{eq}$, build-up times $t_b$
and production rates $R_a$ (see text) versus gas pressure $P$ and laser
power $\cal P_L$ for the data in Fig.\ref{fig:bilan}
and other published data. Results in parenthesis correspond
to strong discharges.}
\label{table:R_a}
\begin{center}
\begin{tabular}{|c|c|c|l|l|l|r@{.}l|} \hline
Ref        &$P$~{\small(mbar)}  & ${\cal P}_{L}$~{\small(W)}
&\multicolumn{1}{c|}{$M_{eq}$~{\small($\%$)}}
&\multicolumn{1}{c|}{$t_b$~{\small(s)}}
&\multicolumn{1}{c|}{$T_1$~{\small(s)}}
&\multicolumn{2}{c|}{R$_a$~{\small(mbar/s)}}\\
\hline
this work     &1.33    &2.0    &80 (60) &67 (20) &300 (60)&0&016  (0.039) \\
\hline
this work     &8       &0.5    &67      &70      &2600    &0&076      \\
\hline
this work     &32      &0.5    &44 (32) &88 (26) &1600 (325)&0&159  (0.401) \\
\hline
this work     &67      &0.5    &24 (20) &84 (33) &1300 (700)&0&191  (0.405) \\
\hline \hline
\cite{Stoltz} &1       &0.05   &50 (40) &40 (9)  &270 (40)&0&013 (0.047) \\
\hline
\cite{Leduc} &1.33    &1.1    &56 (39) &11 (2)  &400 (10)&0&066 (0.266) \\
\hline
\cite{Gentile}&1.33    &4.5    &78 (45) &6.5(0.3)&900 (15)&0&160  (2)   \\
\hline
\end{tabular}
\end{center}
\end{table}
At low pressure, production rates at high field are lower than those
obtained with low-field optical pumping.
Nevertheless, one can take advantage of the weak pressure dependence of
$M_{eq}$ and $t_b$ at $1.5$~T to efficiently perform MEOP at higher pressure.
By increasing the pressure from $1.33$ to $32$~mbar, 
a factor of $10$ in $R_a$ is gained and good production rates are recovered.
For instance, gas in a $250$~cc cell at $32$~mbar can be polarized
at $40\%$ within $3$~minutes. This amount of gas is suitable for small
animal lung imaging after compression to
atmospheric pressure.
For human lung MRI, considerable scaling-up or accumulation of polarized
gas remains necessary. However, optical pumping
around $50$~mbar would considerably simplify the compression stage
by reducing the compression ratio from 1:1000 down to 1:20.

An intrinsic advantage of the high-field MEOP scheme
is that, due to the large Zeeman splittings in the $2^3$S-$2^3$P transition,
the magnetic sublevels involved in optical pumping are selected by
the frequency of the light, and not only by its polarization. 
High-field MEOP is therefore extremely robust
against polarization impurity of the pumping light. This is a crucial
issue for massive production of polarized $^3$He using
high laser power, since imperfect light polarization can severely
limit achieved polarizations at low field~\cite{Leduc}.
\section{Perspectives}
The nuclear polarization improvement observed at
$1.5$~T for high pressures
is plausibly due to the inhibition by hyperfine decoupling of relaxation
channels in atomic and/or molecular states in the plasma, 
as suggested by preliminary results at $0.1$~T~\cite{Cracovie}.
Further experiments at different magnetic field intensities are planned
to confirm this hypothesis. In this perspective, the present study provides 
a first set of data showing that, in spite of
the large hyperfine decoupling in the $2^3$S state, MEOP at high 
field~({\it i})~still yields high nuclear polarizations at low
pressures and~({\it ii})~extends the domain of its applicability to higher
pressures, providing fair polarizations and high production rates.  
From a practical point of view, and in the perspective of a large scale 
medical use of polarized gases,
the development of a $^3$He polarizer operating at $1.5$~T (a widely
 used magnetic field in MRI), and at tens of mbar (for simplified
 compression) could be an attractive choice.

\end{document}